\documentclass[journal]{IEEEtran}
\usepackage{blindtext}
\usepackage[pdftex]{graphicx}

\usepackage{cite}

\ifCLASSINFOpdf
  \usepackage[pdftex]{graphicx}
  \graphicspath{{Figures/}}
\else
\fi
\usepackage[cmex10]{amsmath}
\usepackage{amssymb}
\usepackage{algorithmic}

\usepackage{array}

\usepackage[tight,footnotesize]{subfigure}

\usepackage[font=footnotesize, caption=false]{subfig}
\usepackage{url}

\usepackage{subfiles}
\usepackage{amsfonts}

\usepackage{siunitx}

\usepackage{amsthm}
\theoremstyle{definition}

\usepackage{todonotes}
\usepackage{dirtytalk}
\usepackage{mathtools}
\usepackage[noabbrev, nameinlink, capitalise]{cleveref}
\crefname{paragraph}{Paragraph \nameref}{Paragraphs \nameref}
\Crefname{paragraph}{Paragraph \nameref}{Paragraphs \nameref}

\usepackage{xcolor}
\newcommand{\red}[1]{\textcolor{red}{#1}}

\begin{document}
%
\title{Lippmann Photography: \\ A Signal Processing Perspective}
%
%
%

\author{Gilles~Baechler\(^{*\ddagger}\),
        Michalina~Pacholska\(^{*\ddagger}\),~\IEEEmembership{Member,~IEEE,}
        Arnaud~Latty\(^\dagger\),
        and~Adam~Scholefield\(^\ddagger\),~\IEEEmembership{Member,~IEEE},
        Martin~Vetterli\(^\dagger\),~\IEEEmembership{Fellow,~IEEE,}
\thanks{\(^\dagger\) These authors are with the School of Computer and Communication Sciences, Ecole Polytechnique F\'{e}d\'{e}rale de Lausanne (EPFL), CH-1015 Lausanne, Switzerland. This research was supported by a National Science Foundation grant --- Inverse Problems Regularized by Sparsity --- no. 20FP21151073, and ERC Advanced Grant --- Support for Frontier Research --- SPARSAM no. 247006.}
\thanks{\(^*\) These authors contributed equally. Detailed author contributions: G.B, M.V, and A.S designed research; G.B, A.L, M.P, and A.S performed research; G.B and A.L performed experiments; M.P and G.B ran simulations and wrote the paper; G.B analyzed data.}
\thanks{\(^\ddagger\) These authors contributed while with the School of Computer and Communication Sciences, Ecole Polytechnique F\'{e}d\'{e}rale de Lausanne (EPFL). G.B. is with Google, M.P. is with Improbable and A.S. is with We Play Sport.}
}

\maketitle

\begin{abstract}

Lippmann (or interferential) photography is the first and only analog photography method that can capture the full color spectrum of a scene in a single take. 
This technique, invented more than a hundred years ago, records the colors by creating interference patterns inside the photosensitive plate. 
Lippmann photography provides a great opportunity to demonstrate several fundamental concepts in signal processing. Conversely, a signal processing perspective enables us to shed new light on the technique. 

In our previous work \cite{baechler2021shedding}, we analyzed the spectra of historical Lippmann plates using our own mathematical model.
In this paper, we provide the derivation of this model and validate it experimentally. We highlight new behaviors whose explanations were ignored by physicists to date. In particular, we show that the spectra generated by Lippmann plates are in fact distorted versions of the original spectra. We also show that these distortions are influenced by the thickness of the plate and the reflection coefficient of the reflective medium used in the capture of the photographs. We verify our model with extensive experiments on our own Lippmann photographs.

\end{abstract}



%
\IEEEpeerreviewmaketitle


\section{Introduction}

Lippmann photography is one of the oldest color photographic techniques and the oldest multispectral photographic method. In 1908 Gabriel Lippmann received the Nobel Prize in physics both for the invention of this technique and for the physical explanation of the process \cite{NobelPrize1908Physics}. Yet, until now, this photographic method was not fully understood.

\begin{figure}[!tb]
\centering
	\includegraphics[width=1\linewidth]{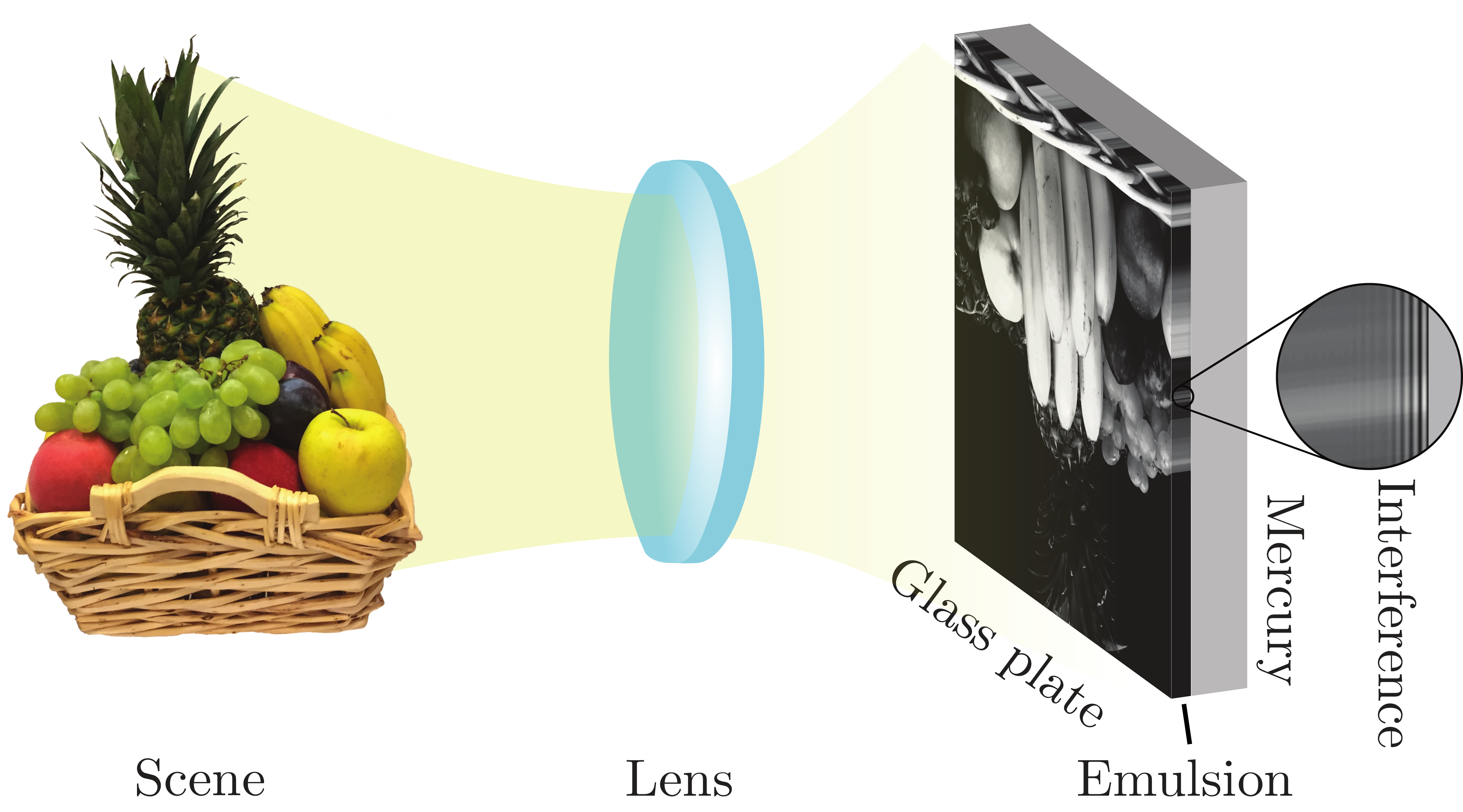}
\caption[]{Principle of Lippmann photography: the scene on the left is focused with a photographic lens on the plane of the emulsion. Behind the emulsion is a mercury bath; when the light wave hits the mercury layer, it reflects, and interferes with itself. Its spatially varying intensity creates patterns that are captured inside the depth of the  emulsion (see the close-up).
\label{fig:lippmann_photography}
}
\end{figure}

At a high level, the Lippmann process works by capturing an interference pattern in a 
photosensitive emulsion. As shown in \cref{fig:lippmann_photography}, the photographic plate consists of a
light-sensitive emulsion on a sheet of glass. A mirror is created at the surface of the emulsion, traditionally by putting the emulsion in contact with liquid mercury. The plate is oriented such that the light from the scene passes through the glass then through the emulsion, before reflecting back from the mirror. 
The reflected light interferes with the incoming light and the  resulting interference pattern exposes the emulsion differently at different depths. Each point of the scene is focused by usual camera optics onto a point of the mirror, so the interference pattern is generally specific to each point ("pixel") of the image.

Once the plate is exposed, it is removed from the liquid mercury and 
processed via standard photographic development techniques~\cite[Chapter~4]{Bjelkhagen1993}. 

\begin{figure}[!tb]
\centering
	\includegraphics[width=1\linewidth]{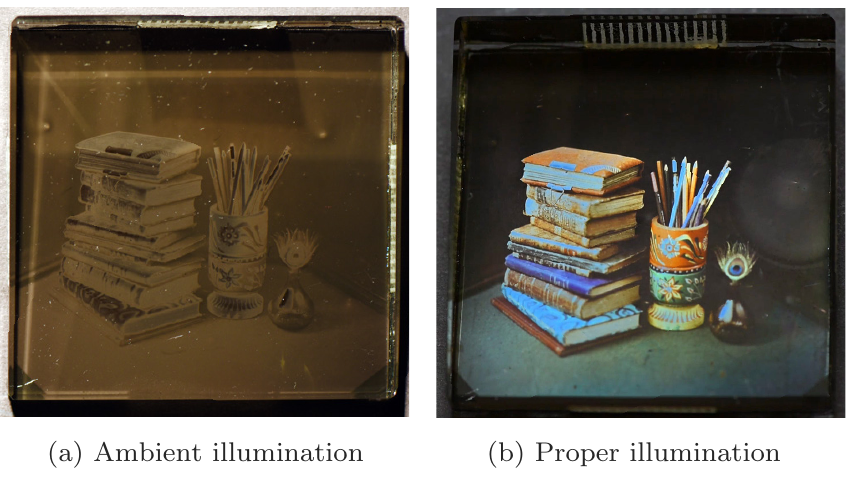}
\caption[]{A Lippmann plate, which has been created by photographer Filipe Alves with a home-made albumen emulsion: under (a) ambient illumination and (b) directed light whose incoming direction is the mirror of the viewing direction with respect to the surface normal.
\label{fig:lippmann_photos} 
}
\end{figure}

The recorded image becomes visible when the plate is illuminated by a white light source perpendicular to the plate (parallel to \(z\)-axis), see \cref{fig:lippmann_photos}. The incoming light is scattered by the tiny silver particles that are distributed throughout the emulsion.  The light waves reflected at different depths interfere with each other, and depending on the recorded interference patterns, some reflected colors interfere constructively and others do so destructively. 

\begin{figure}[!tb]
\centering
\begin{tikzpicture}
\node[] at (0,0) {\includegraphics[width=0.95\linewidth]{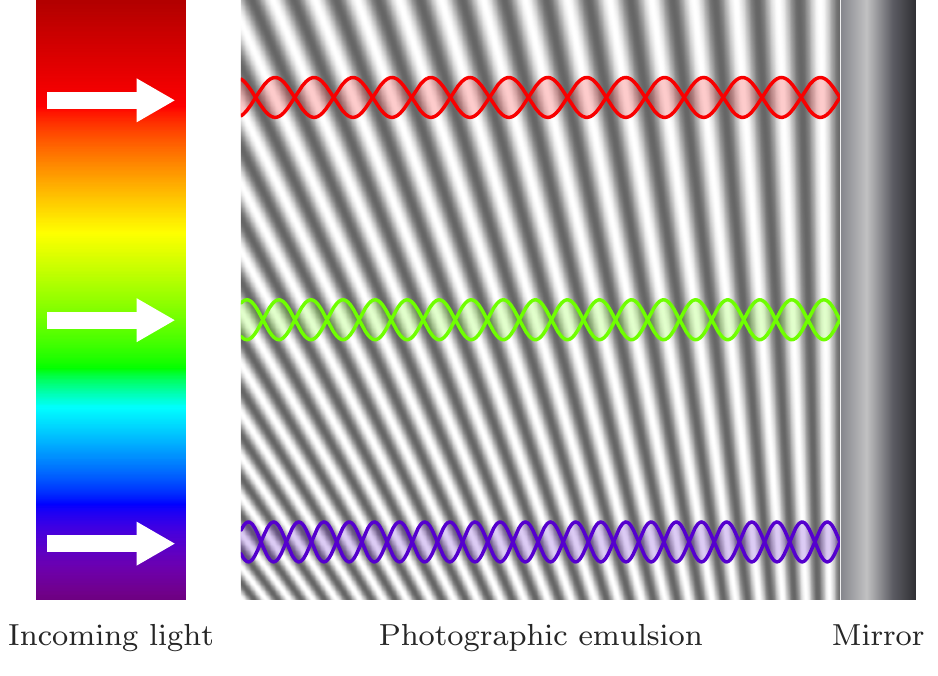}};
\draw [-stealth](4.2,-1) -- (4.2,1);
\node[rotate=90] at (4.5,0) {\(x\)};
\node[] at (0,3.5) {\(z\)};
\draw [-stealth](1,3.2) -- (-1,3.2);
\end{tikzpicture}
\caption[]{Schematics of a Lippmann photograph of a rainbow created by dispersing white light through a prism: at each \(x\) position, a monochromatic wave enter the plate and the interference with its reflection creates standing waves in the photographic emulsion. Different colors produce different standing waves and therefore different patterns in the emulsion. 
\label{fig:lippmann_mono_plate} 
}
\end{figure}
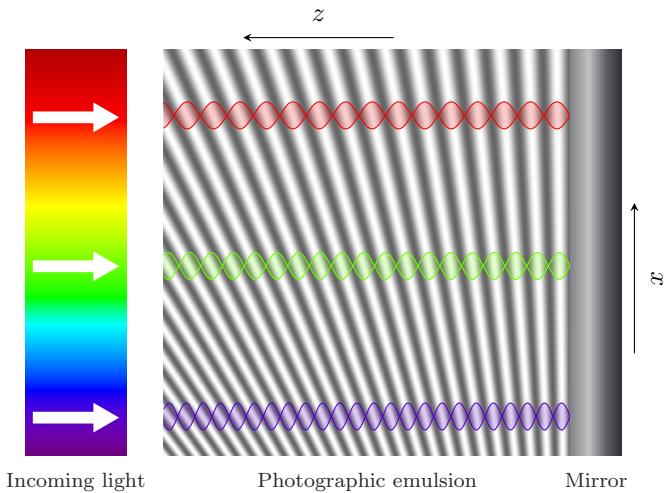

For example, in the case of monochromatic light, the interference pattern takes the form of cosines, whose spatial period is \(\lambda/(2n)\), where \(\lambda\) is the wavelength of the light wave in vacuum and \(n\) is the index of refraction of the material. The resulting silver grains are arranged in layers parallel to the mirror inside the plate, and can be thought of as thin films. \Cref{fig:lippmann_mono_plate} provides a schematic view of a plate illuminated with different monochromatic colors.

The strongest reflection from thin films \(\lambda/(2n)\)-apart is obtained for waves of wavelength \(\lambda\) (and their harmonics).
Therefore, we can expect the Lippmann plate to
reflect light of the same color that it was
illuminated with during recording, at least for
monochromatic waves.

A persistent misconception in the literature is that Lippmann plates perfectly store any (polychromatic) spectrum, at least from a theoretical point of view; that is, assuming a plate of infinite thickness, uniform development and a re-illumination with a flat spectrum. 
In fact, Lippmann himself made such a claim in his 1894 paper using the premises of Fourier analysis~\cite{Lippmann1894}:\\

\say{en d'autres termes, la couleur de l'image est la m\^{e}me que celle de l'objet.}\footnote{\say{in other words, the color of the image is the same as that of the object.} (our translation)}\\

\noindent More recently, H. Bjelkhagen, an expert in Lippmann photography and holography, wrote~\cite{Bjelkhagen2013}:\\

\say{The light is reflected from these fringes, creating precisely the colors that correspond to the original ones that had produced them during the recording.}\\

In recent work \cite{baechler2021shedding}, we have shown that for polychromatic spectra, the reflected spectrum is not exactly the same as the recorded spectra. We proposed a mathematical model of the Lippmann plates. We then used it to analyze spectra reflected from historical plates and propose how to correct arising distortions. 

In this work, we derive and explain our model in more detail as well as verify experimentally the predicted distortions. Here, we take a signal processing view of the Lippmann process, whereas \cite{baechler2021shedding} is aimed at more general audience and hence lacks mathematical detail. In \cref{sec:mathematical} we describe and model mathematically all three stages of the Lippmann process: recording, development and viewing. Then, in \cref{sec:verification} we analyze the predictions of our model and confirm them experimentally. We show that the reflected spectrum contains oscillations not present in the recorded spectrum, and that the resulting color is skewed towards red (if the plate is recorded with mercury). Many of our observations have never been documented even in modern studies~\cite{Nareid1988, Nareid1991, Fournier1998, Bjelkhagen2013} (\cref{sec:sota}). The results presented are outcomes of the theses of the first two authors \cite{baechler2018sampling, pacholska2021sampling}.

\subsection{Previous work}
\label{sec:sota}

Gabriel Lippmann was not the only person to have worked on Lippmann photography. The Lumi{\`e}re brothers \cite{Lumiere1893, Lumiere1893a}, Wiener~\cite{Wiener1899}, Roth\'{e}~\cite{Rothe1904} and Ives~\cite{Ives1908}, analyzed and improved Lippmann plates.
 Neuhauss~\cite{Neuhauss1898}, 
Lehmann~\cite{lehmann1906}, and Cajal~\cite{Cajal1906,Cajal1907} observed the interference patterns of Lippmann plates under optical microscopes by swelling the gelatin. Wiener demonstrated the interference pattern by recording a slice through it at a very small angle \cite{Wiener1890}.

Despite all the interest in Lippmann photography, it never became practical. It was not consistent enough to be commercialized 
\cite{Mitchell2010}. Due to the small size of the silver grains, the exposure times were long. Moreover, the Lippmann plates were impossible to copy.

More recently, Lippmann photography received renewed attention, partly due to developments in holography \cite{GABOR1948, gabor1949microscopy}. Like Lippmann photography, holography also records the interference of light. However, holography uses a reference wave --- not the 
reflected wave --- to create interference. This means that holography can 
record the phase of light, but also that it has to be recorded with coherent light. Although the photographic processes differ, holographic plates can be easily adapted for the Lippmann process.
Therefore, the industrial production of holography plates enabled more people to experiment with Lippmann photography \cite{bjelkhagen2001lippmann}.
                                                 
Since the 1980s, there have been a number of works analyzing Lippmann photographs, but none of them combined detailed mathematical modeling with experimental results.
Phillips et al.~\cite{Phillips1984, Phillips1985} precisely described the 
scattering of the silver grains, as well as their size limitations. 
Fournier et al.~\cite{Fournier1994, Fournier1998} studied the 
creation of extremely fine-grain emulsions and analyzed the structure and 
spectrum reproduction of historical and contemporary photographic plates. 
Nareid and Pederson~\cite{Nareid1988, Nareid1991} introduced a mathematical  model based on local changes in the refractive index of the plate. Alschuler described observations from creating his own Lippmann photographs \cite{Alschuler2002}.
Last but not least, Bjelkhagen studied many practical aspects of the technique, such as recording with holographic plates and the use of the method for security purposes~\cite{Bjelkhagen1997, Bjelkhagen2003, Bjelkhagen2002, Bjelkhagen2013}.

\begin{figure*}[!tb]
\centering
\begin{tikzpicture}
\node[inner sep=0pt] (pipeline) at (0,0)
    {\includegraphics[width=1\linewidth]{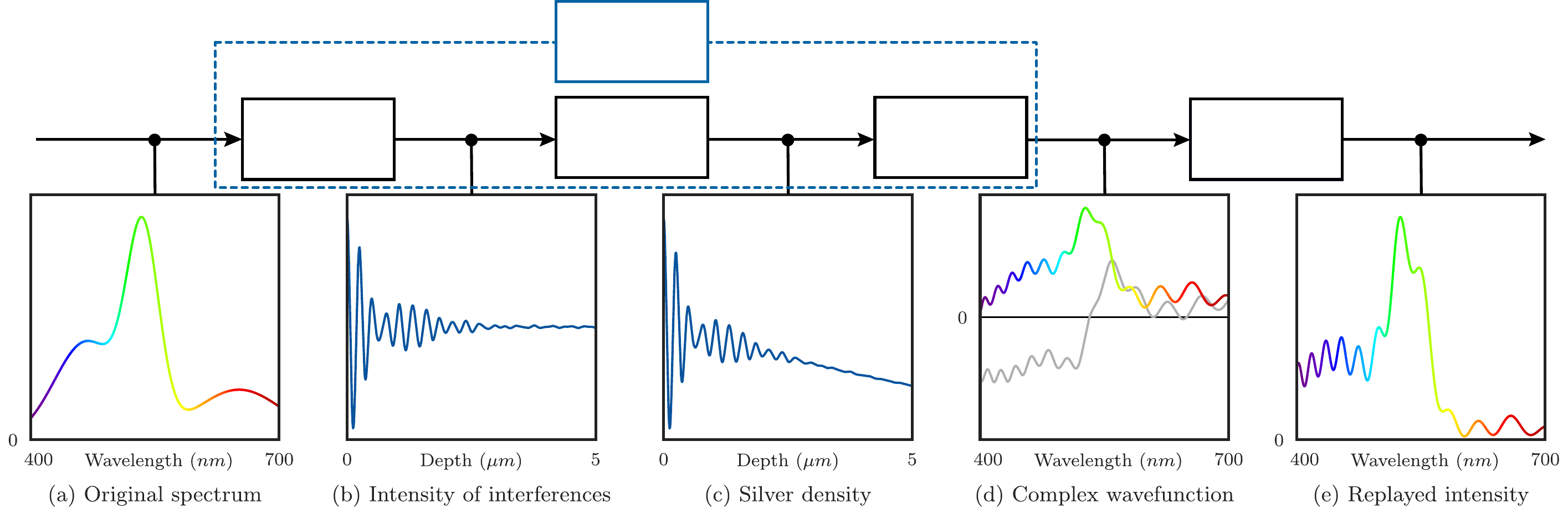}};
\node[inner sep=0pt] (P) at (-7.4,1.7) {\(P(\omega)\)};
\node[inner sep=0pt] (I) at (-3.6,1.7) {\(I(z)\)};
\node[inner sep=0pt] (rho) at (0,1.7) {\(\rho(z)\)};
\node[inner sep=0pt] (Ur) at (3.8,1.7) {\(U_r(\omega',t)\)};
\node[inner sep=0pt] (Pr) at (7.5,1.7) {\(P_r(\omega')\)};
\node[inner sep=0pt] (rec) at (-5.45,1.4) {\small recording};
\node[inner sep=0pt] (dev) at (-1.8,1.4) {\small development};
\node[inner sep=0pt] (view) at (2,1.4) {\small viewing};
\node[inner sep=0pt] (pow) at (5.8,1.4) {\(|\cdot|^2\)};
\node[inner sep=0pt] (h) at (-1.8,2.55) {\(H(\omega, \omega')\)};
\end{tikzpicture}
\caption[]{Lippmann photography pipeline: (a) the original spectrum \(P(\omega)\) undergoes a linear transform and (b) its interference pattern \(I(z)\) is captured inside the plate; the plate is then developed and (c) the silver density \(\rho(z)\) at depth \(z\) is proportional to \(I(z)^2\), modulated by the decay \(e^{-2\tau z}\); when the plate is illuminated, the reflected wave (d) $U_{r}(\omega', t)$ is similar to filtering $P(\omega)$ with \(H(\omega, \omega')\), the colored curve represents its real part and the gray curve its imaginary part; finally, the observed spectrum \(P_r(\omega')\) is given by the time-averaged intensity of $U_{r}(\omega', t)$.}
\label{fig:pipeline} 
\end{figure*}
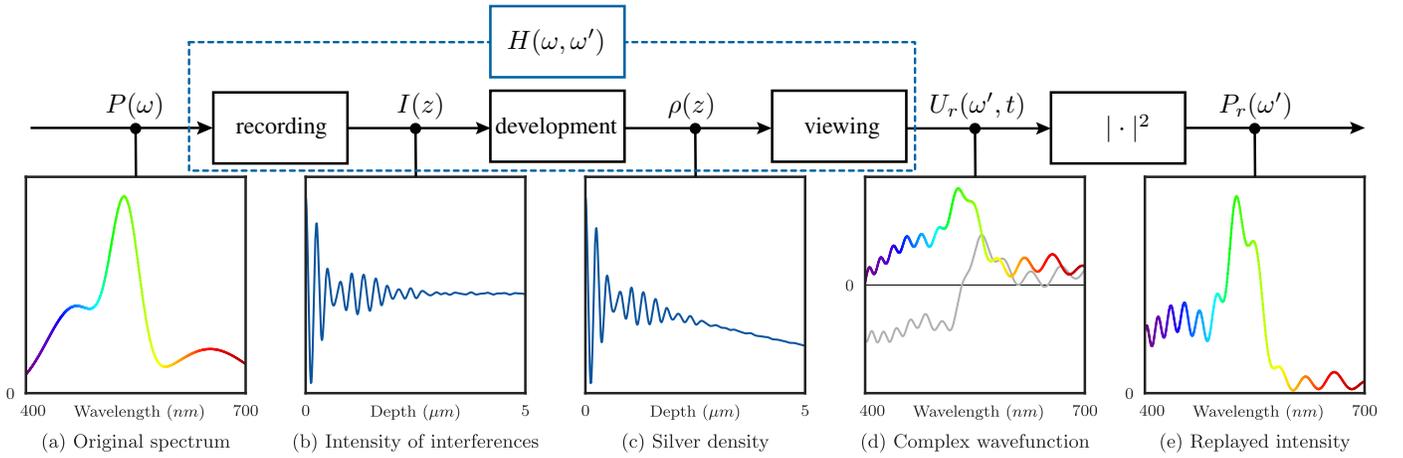

\section{Mathematical Model}
\label{sec:mathematical}

In this section, we propose a mathematical model for the Lippmann process.
We model the Lippmann process as a signal-processing pipeline, as illustrated in \cref{fig:pipeline}.
We interpret the exposure as applying an analysis operator, the effects of
development as filtering, and the viewing as applying a synthesis operator and
taking the squared magnitude.

Throughout this paper, we model light fields as plane waves.
We assume that all waves travel in the direction perpendicular to the surface
of the plate. Our assumption is not realistic because a planar wave
has the same color and intensity everywhere on the plane perpendicular to the direction of the propagation. Nevertheless, such waves are a
reasonable approximation if the scale of variation in color or phase is much
larger than the wavelength of light, which is true for artistic photography. This assumption also means that we model a single \say{pixel}, i.e., a small region where the planar assumption is valid. 

A heterogeneous $1$D wave $U(z,t)$, traveling in the direction of 
the \(z\)-axis, can be expressed as an integral over individual sine waves, each having an amplitude $A(\omega)$:
\begin{align}
U(z,t) &= \int_{0 }^{\infty} A(\omega) e^{j \left(\omega t \mp \frac{\omega z}{c} \right)} d\omega,
\end{align}
where \(\omega\) is the angular frequency, $c$ is the speed of light in the medium, $z$ and $t$ are, respectively, the spatial and temporal coordinates, and the sign \(\mp\) depends on the direction of the propagation of the wave.
For visible light, we can measure only the power spectral density of \(U\): 
\begin{align}
P(z, \omega) = P(\omega) = \left|A(\omega)\right|^2.
\end{align}

\subsection{Analysis: Recording a Lippmann plate}
\label{sec:analysis}

From a signal processing perspective, Lippmann photography works by recording the Fourier cosine transform of the light spectrum in the thickness of a photographic emulsion. The emulsion is usually made of gelatin and contains extremely fine light-sensitive silver-halide grains as well as color dyes that make it sensitive to all visible wavelengths.\footnote{Since Lippmann recorded his first photographs, other materials such as photopolymers~\cite{Bjelkhagen1997}
have been used to record the interference field. In holography, people use photoresist and dichromated gelatin~\cite{Saxby2015}.}
During the recording, the emulsion is put in direct contact with a mirror. This way, the light reflected from the mirror creates interference patterns inside the emulsion. In this section, we model this process mathematically. 

\begin{figure}
    \centering
    \begin{tikzpicture}
    \node[inner sep=0pt] (pipeline) at (0,0){
    \includegraphics[width=\linewidth]{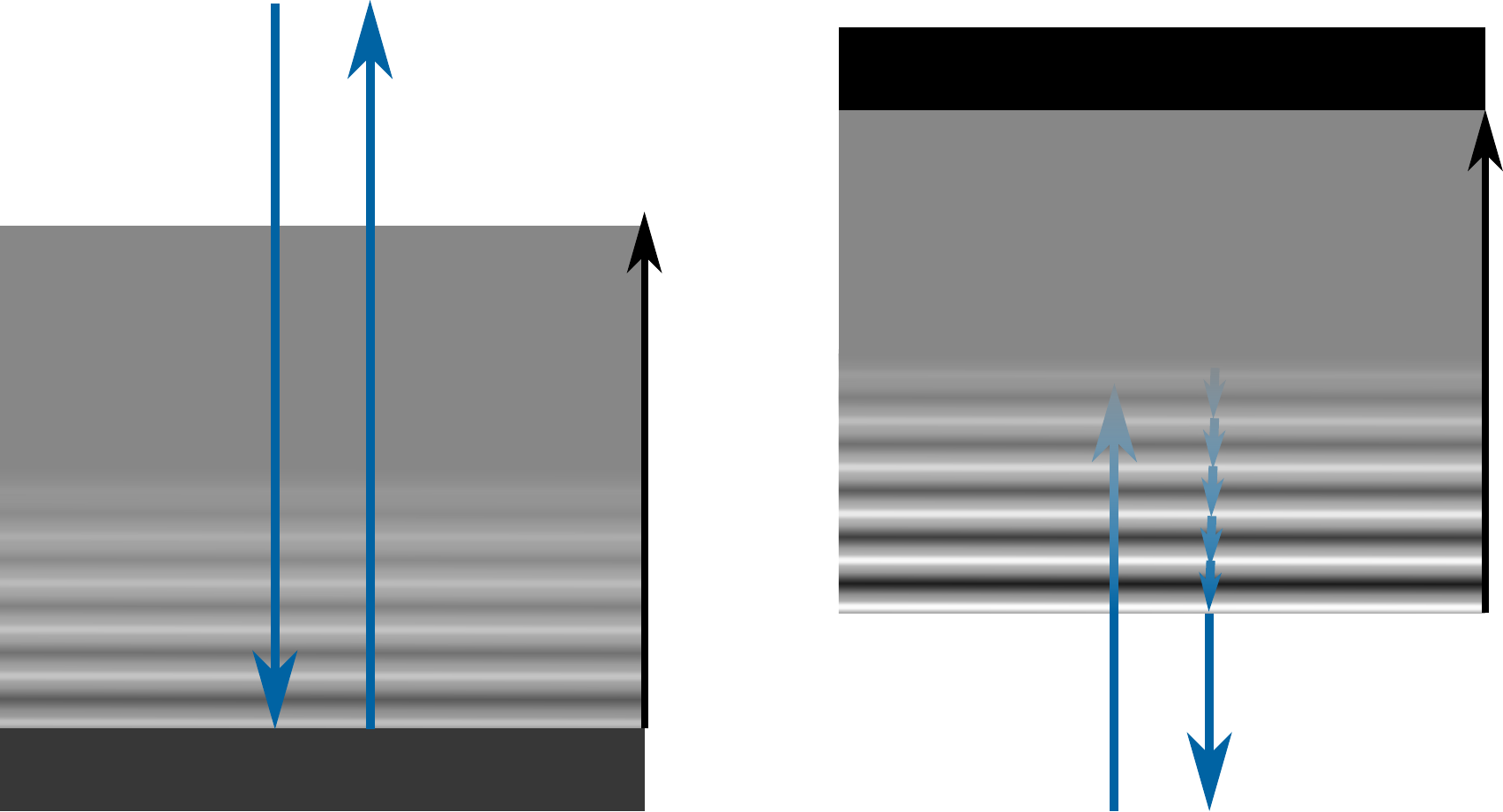}};
    \node[inner sep=0pt] (z1) at (-0.4, 0.6) {\(z\)};
    \node[inner sep=0pt] (01) at (-0.4, -1.7) {\(0\)};
    \node[inner sep=0pt, text=white] (P) at (-2.5, -2.15)
    {mirror};
    \node[inner sep=0pt, text=white] (P) at (-2.5, -2.8)
    {(a)};
    \node[inner sep=0pt] (z2) at (4.6, 1.3) {\(z\)};
    \node[inner sep=0pt] (02) at (4.6, -0.9) {\(0\)};
    \node[inner sep=0pt, text=white] (P) at (2.4, 1.95)
    {black paint};
    \node[inner sep=0pt, text=white] (P) at (2.4, -2.8)
    {(b)};
    \node[inner sep=0pt] (z2) at (-3.1, 1.5) {\(U\)};
    \node[inner sep=0pt] (02) at (-1.9, 1.5) {\(U'\)};
    \node[inner sep=0pt] (z2) at (3.05, -1.6) {\(U_r\)};
    \node[inner sep=0pt] (02) at (1.75, -1.6) {\(U_i\)};
    \end{tikzpicture}
    \caption{Recording (a) and viewing (b) of the plate in the coordinates used in this section. The origin is at the edge of the plate that is adjacent to the mirror during the recording. During the recording the incoming wave \(U\) travels towards negative \(z\), and during replay the incoming wave \(U_i\) travels towards positive \(z\).}
    \label{fig:coordinates}
\end{figure}
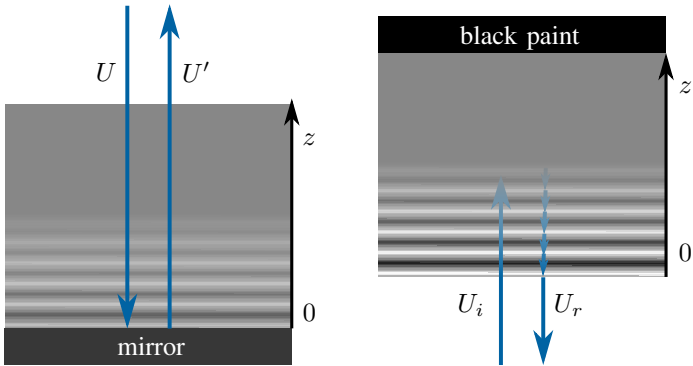

By convention, we set $z=0$ to be the position of the mirror and assume that the incoming wave $U_0$ travels in the direction of negative \(z\). Upon hitting the mirror, the wave is reflected back towards positive \(z\), see \cref{fig:coordinates}.
We define the reflection coefficient of the mirror as $r = \rho e^{j \theta}$, where $\rho$ is the attenuation factor and $\theta$ the phase shift. This leads to the following expression for the reflected wave that travels towards positive \(z\):
\begin{align}
U_1(z,t) &= r \int_{0 }^{\infty} A(\omega) e^{j \left(   \omega t - \frac{\omega z}{c} \right)} d\omega.
\end{align}
The resulting wave is then composed of both the incoming and the reflected fields:
\begin{align}
U_{\text{tot}}(z,t) &= U_0(z,t) + U_1(z,t) \nonumber\\ 
&= \int_{0 }^{\infty} A(\omega)  e^{j \omega t }  \left( e^{j\frac{\omega z}{c}} + re^{-j\frac{\omega z}{c}} \right)  d\omega.
\end{align}
The time-averaged intensity of this wave is therefore given by
\begin{align}
\label{eq:intensity_interferences}
I (z)
&= \int_{0 }^{\infty} |A(\omega)|^2 \left| e^{j\frac{\omega z}{c}} + r e^{-j\frac{\omega z}{c}}  \right|^2 d\omega \nonumber\\
&= \int_{0 }^{\infty} P(\omega) \left( 1 + \rho^2 + 2\rho \cos\left(\frac{2 \omega z}{c } - \theta  \right) \right) d\omega.
\end{align}
The intensity \(I\) varies spatially in $z$ and takes the form of \emph{standing waves}\footnote{Technically speaking, standing waves only form in the monochromatic case. In the polychromatic case, we can think of the shape of the resulting wave as a superposition of standing waves.}, or more generally \emph{partial} standing waves when $\rho < 1$.
Since this is happening inside the depth of a photosensitive plate, these interference patterns expose the photographic emulsion differently at different depths.

In the literature~\cite{Lippmann1894, Bjelkhagen2013, Nareid1991}, perfect reflection is often assumed, that is either $r = -1$ or $r = 1$, depending on whether there is a phase inversion at the interface:
\begin{align}
I_{r = 1}(z)
&= 2\int_{0 }^{\infty} P(\omega) \left( 1 + \cos\left(\frac{2 \omega z}{c }  \right) \right) d\omega,\\
I_{r = -1}(z)
&= 2\int_{0 }^{\infty} P(\omega) \left( 1 - \cos\left(\frac{2 \omega z}{c }  \right) \right) d\omega.
\end{align}
We keep the full model from~\eqref{eq:intensity_interferences} as it allows for a more precise analysis; as we will see later in \cref{sec:spectrum_skewing}, the value of $r$ has a strong influence on the resulting spectrum.

In summary, the interference patterns inside a Lippmann plate are superpositions of partial standing waves, which are described by the sum of a constant and an oscillating term.
The oscillating term 
\begin{align}
\mathcal{F}_{\theta}\{P\}(z) = 2\rho \int_{0 }^{\infty} P(\omega) \cos\left(\frac{2 \omega z}{c } - \theta  \right) d\omega
\end{align}
is in fact the \emph{generalized Fourier cosine transform} of $P$, and it has been shown in \cite{moiseev1998generalized} that it is invertible~\cite{baechler2018sampling}.
Therefore, by analyzing an infinite-depth plate, one could separate the constant and oscillating terms and invert the recording step.

\subsection{Filtering: Development}
\label{sec:brief_development}

From a chemical point of view, exposing the photographic emulsion to light breaks some of the bonds between silver particles and halide ions by reducing the silver. Once free, these metallic silver atoms form tiny specks that combine into a \emph{latent image}. 

In this work, we assume that the relationship between the exposing intensity and resulting metallic silver density is quadratic:
\begin{align}
    \label{eq:silver_density}
    \rho_{\text{lat}}(z) \propto I (z)^2,
\end{align}
where \(\rho_{\text{lat}}\) is the metallic silver density at depth $z$.\red{\footnote{This assumption is an educated guess. The relationship between exposure and (final) silver density, described by the so-called \emph{Hurter and Driffield curve}, is close to a power law in the useful exposure range (which we found empirically for our material and process)~\cite[Chapter~2]{Bjelkhagen1993}. The exponent (development factor or \emph{gamma}) is typically between 1.5 and 2 with the kind of developer we use (pyrogallol-ammonia)~\cite{HurterDriffield1898}, measured on regular photographic emulsions. As \say{Holographic emulsions are of the high-contrast (high-gamma) type}~\cite[Chapter~2]{Bjelkhagen1993}, higher values of gamma cannot be ruled out.}}
As we will see in \cref{sec:full_pipeline}, this assumption results in the linear dependence of the \emph{amplitude} of the reflected wave on the \emph{power} of the incoming wave.
Furthermore, as we show in \cref{sec:verification}, our model accurately predicts experimental results.

After the plate has been exposed, the next step is to develop it. 
Loosely speaking, the role of development is to increase the concentration of metallic silver in the exposed areas. It is important to note that the development is not uniform throughout the thickness of the plate and deeper layers are generally less developed than the top layers. 
For simplicity, we model the effect of the developer as a multiplicative factor that decays exponentially with the depth of the plate, with the decay rate \(2\tau\). The silver density \(\rho_{\text{dev}}(z)\) after the development is then
\begin{align}
\label{eq:development}
\rho_{\text{dev}}(z) =  e^{-2\tau z} \rho_{\text{lat}}(z) \propto \left(e^{-\tau z} I (z)\right)^2.
\end{align}
Here, we have included the development inside the power because it will be useful in the combined model of recording, development, and viewing.

After the development bath, a fixing process is optionally applied to wash out dyes, halide crystals and silver halides~\cite{Saxby2015}, so that the plate is no longer sensitive to light.

\subsection{Synthesis: Viewing a Lippmann Plate}
\label{sec:synthesis}

To visualize a developed plate, we remove the mirror and illuminate the plate from the other side (see \cref{fig:coordinates}) with a light source with power spectral density \(P_i(\omega)\).
The incoming light is scattered by the silver particles that are distributed throughout the depth of the emulsion. From a higher level, the plate can be thought of as consisting of elementary partially reflective mirrors, whose reflectance is proportional to the metallic silver density at that point. 

\begin{figure}[!tb]
\centering
	\includegraphics[width = 1\linewidth]{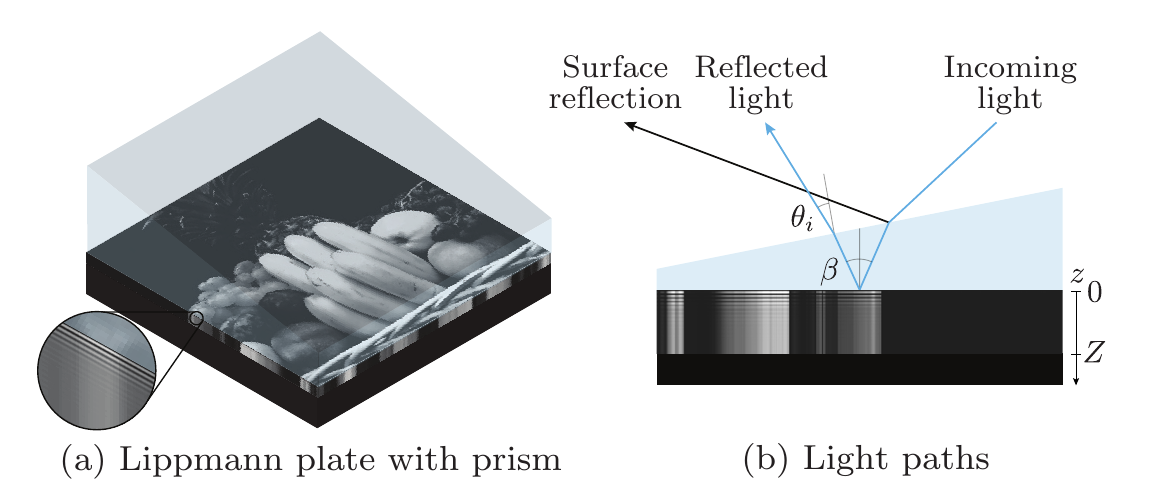}
\caption[]{Lippmann plate prepared for visualization after development; (a) the glass plate has been painted in black to minimize reflections and a prism has been mounted on the side that was in contact with the mirror; (b) the light rays reflecting at the surface of the prism do not coincide with the rays reflected from the plate.
\label{fig:lippmann_replay} 
}
\end{figure}

Note that here we strictly describe the interaction of the light \emph{inside} the plate, but do not take into account the reflection of the light at the surface of the plate. This reflection is generally strong compared to the optical wave reflected from within the plate. To cope with this issue, a prism with a refractive index close to that of gelatin is usually attached to the top of the plate; this prevents the surface reflection from mixing with the internal reflections (see \cref{fig:lippmann_replay}).

\begin{figure}
    \centering
    \begin{tikzpicture}
    \node[inner sep=0pt] (pipeline) at (0,0){
    \includegraphics[width=\linewidth]{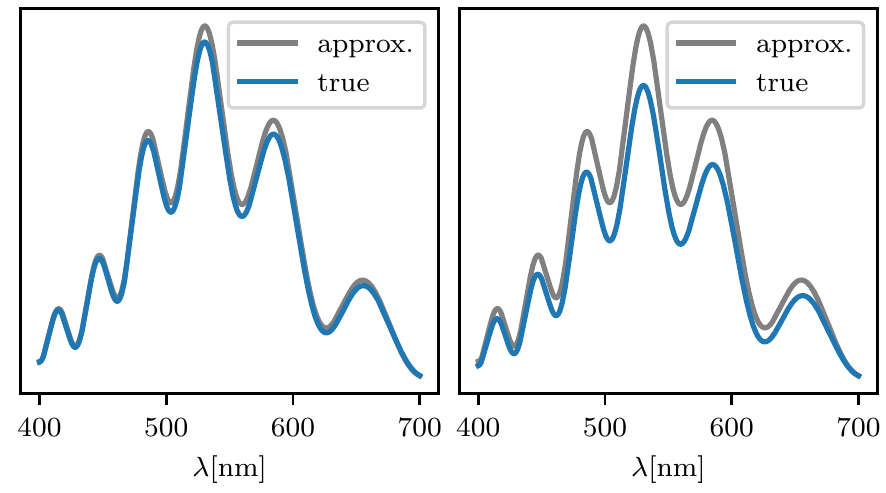}};
    \node[] (a) at (-2.1, -2.7) {(a)};
    \node[] (b) at (2.3, -2.7) {(b)};
    \end{tikzpicture}
    \caption{Comparison of the approximated model of reflection \eqref{eq:reflected_power} (gray) with the numerical simulations using the transfer-matrix method (blue). The plate was illuminated with white spectrum. Left: for small values of reflectivity \(\rho_{\text{dev}}\) the approximation is accurate. Right: for \(10\) times higher values of reflectivity \(\rho_{\text{dev}}\) the approximation overestimates the reflected power, but the overall shape of the spectrum is preserved.}
    \label{fig:approximation}
\end{figure}

Like Lippmann, we consider only the first-order reflections, i.e., we neglect all inter-reflections. In addition, we assume that the amount of light that hits each layer is always the same regardless of the depth. This is clearly not the case in reality: deeper layers cannot receive as much light as shallow layers, since some light is already reflected by these shallow layers. Nevertheless, these assumptions are reasonable when the amount of reflected light remains relatively small, see \cref{fig:approximation}. More importantly, they enable us to derive a closed-form solution of the Lippmann procedure, described in \cref{sec:full_pipeline}.

As the light transport can be assumed linear for the energies of interest, we can consider each frequency component separately and superpose the results.
Let \(U_i\) be a  monochromatic wave with frequency \(\omega'\) that illuminates the plate.
The \emph{power} of the wave \(U_r(z)\) reflected from the slice of the plate at depth \(z\) is proportional both to the silver density inside the slice \(\rho_{\text{dev}}(z)\) and the power of the incoming wave \(U_i\):
\begin{equation*}
    P_r(z,\omega') \propto \rho_{\text{dev}}(z) P_i(\omega').
\end{equation*}
Therefore, the amplitude of the reflected wave is proportional to the square root of the density \(\rho\):
\begin{equation*}
    A_r(z,\omega') \propto \sqrt{\rho_{\text{dev}}(z)} A_i(\omega').
\end{equation*}

Each component  $dU_i(\omega')$ of the incoming wave $U_i(\omega')$ reflected at depth $z$ travels a round trip distance of $2z$ between the entrance of the plate and the depth $z$: this corresponds to a phase shift of $2 \omega' z/c$. 

In summary, the wave reflected at depth $z$ and measured at the entrance of the plate is given by
\begin{align}
\label{eq:partial_reflected_waves}
dU_{r}(z, \omega', t) \propto  \sqrt{\rho_{\text{dev}}(z)} e^{ j (\omega'  t - \frac{2\omega' z}{c})} A_i(\omega') dz.
\end{align}
The total reflected wave $U_{r}$ is the integral over the whole thickness of the emulsion of the partial reflected waves:
\begin{align}
\label{eq:total_reflected_wavefunction}
U_{r}(\omega', t) &= \int_{0}^{Z} dU_{r}(z, \omega', t) \nonumber\\
& \propto A_i(\omega') e^{  j \omega' t } \int_{0}^{Z} \sqrt{\rho_{\text{dev}}(z)} e^{ -j\frac{2\omega' z}{c}} dz,
\end{align}
where \(Z\) is the thickness of the plate.

 In practice, we can measure only the power spectral density of the reflected wave
\begin{equation}
\label{eq:reflected_power}
    P_r(\omega') \propto P_i(\omega') \left| \int_{0}^{Z} \sqrt{\rho_{\text{dev}}(z)} e^{ -j\frac{2\omega' z}{c}} dz \right|^2.
\end{equation}

In the remaining part of this paper, we assume that the plate is illuminated with an equienergetic white source, i.e., \(P_i(\omega)\) is constant for the frequencies of interest. Visualisation with any light source can be calculated from this source by scaling the reflected frequencies proportionally to their power in the light source.

\subsection{Full pipeline}
\label{sec:full_pipeline}
In this section we combine the equations describing recording \eqref{eq:intensity_interferences}, development \eqref{eq:development} and viewing \eqref{eq:reflected_power} and obtain the relation between recorded and reflected spectrum.

For simplicity, we combine all proportionality factors, including the power of the incoming light, into one constant \(\epsilon\). 
Using~\eqref{eq:development} and~\eqref{eq:reflected_power}, we obtain:
\begin{equation*}
        P_r(\omega') = \epsilon \left|\int_{0}^{Z} e^{-\tau z} I(z) e^{ -j\frac{2\omega' z}{c}} dz \right|^2.
\end{equation*}
Combining it with \eqref{eq:intensity_interferences}, we obtain the squared integral operator that maps recorded spectrum to reflected spectrum:
\begin{align}
\label{eq:reflected_wave_function}
P_r(\omega')
&=  \epsilon \left| \int_0^\infty P(\omega) H(\omega, \omega') d\omega \right|^2,
\end{align}
where we define the function \(H(\omega, \omega')\) as
\begin{align}
\label{eq:filter_h}
\int_0^Z \left( 1 + \rho^2 + 2\rho \cos\left(\frac{2\omega z}{c} - \theta  \right) \right) e^{ -j\frac{2\omega' z}{c}} e^{-\tau z} dz.
\end{align}

The formulation of \eqref{eq:reflected_wave_function} resembles the
formulation of integral transforms in kernel-reproducing Hilbert spaces (KRHS). However, our operator \(H\) is not symmetric, hence we cannot use the core results from KRHS that are based on spectral decomposition of the integral operator. 

\begin{figure}[!tb]
\centering
	\includegraphics[width=1\linewidth]{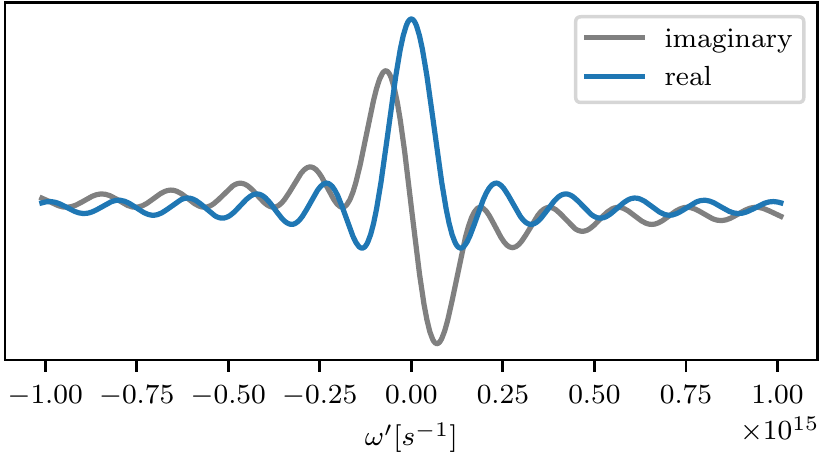}
\caption[]{The filter \(S(\omega')\) for \(Z=5~\si{ \mu} m\) and \(\tau=0\). The real part is represented by the blue line and the imaginary part by the gray line.
\label{fig:h} 
}
\end{figure}

At the end of this section, we show that the kernel \(H\) can be also expressed using a filter \(S\):
\begin{gather}
\label{eq:relationship_h_g}
H(\omega, \omega') = 
r^{*} S(\omega'\!\!-\!\omega) +  (1\!+\!\rho^2)S(\omega') + r S(\omega'\!\!+\!\omega), \\ \nonumber
S(\omega) = \frac{1}{2j\omega/c + \tau}\left( 1- e^{-(\tau + 2j \omega/c)Z } \right).
\end{gather}
Before we derive \(S\), let us make some observations. First, an example of \(S(\omega)\) is shown in \cref{fig:h,fig:h_dev}. It has sinc-like shape and decays like \(1/\omega\). Therefore, if we assume that the plate is sensitive only to visible light and that we measure only visible light, then the third term \(r S(\omega'\!\!+\!\omega)\) is negligible.

The remaining two terms have significant effects on the spectrum. The first term \(r^{*} S(\omega'\!\!-\!\omega)\), which we call \emph{filtering term}, is a low-pass filter. The second term \((1\!+\!\rho^2)S(\omega') \) introduces distortions which we analyze in \cref{sec:verification}. We call this term \emph{constant term}, as it does not depend on the shape of the recorded spectrum. The result of integration of this term with the power spectrum is just multiplication of this term by the total energy of \(P\): \(\int_0^\infty P(\omega)d\omega\).

\begin{proof}[Proof of \eqref{eq:relationship_h_g}]
Let us introduce two new variables, \(\xi = 2\omega/c\) and \(\xi' = 2\omega'/c\). By expanding the cosine into a sum of two exponentials, we can write \eqref{eq:filter_h} as a sum of three terms
\begin{align*}
\int_0^Z \left( \underbrace{1 + \rho^2}_{\text{(i)}} + \underbrace{\rho e^{-j \theta}e^{j\xi z}}_{\text{(ii)}} + \underbrace{\rho e^{j \theta}e^{-j\xi z}}_{\text{(iii)}}  \right)  e^{ -(j\xi'+\tau )z}dz.
\end{align*}
The first term is constant, so its integral will be equal to \( \left( 1 + \rho^2 \right)  s_{(i)}\), where 
\begin{align*}
 s_{(i)} &= \int_0^Z  e^{ -(j\xi'+\tau )z}dz \\
&=  \left[ \frac{-1}{j\xi'+\tau} e^{ -(j\xi'+\tau )}\right]_{0}^{Z} \\
&=  \frac{1}{j\xi'+\tau} \left(1 - e^{ -(j\xi'+\tau )} \right).
\end{align*}
By reversing the change of variables, we obtain that 
\( s_{(i)} = S(\omega')\).
Similarly, the second term can be written as:
\begin{align*}
 \rho e^{-j \theta} \int_0^Z  e^{ -(j(\xi' - \xi)+\tau )z}dz.
\end{align*}
The integral is exactly the same as \(s_{(i)}\), except \(\xi'\) is replaced by \(\xi'-\xi\). Therefore, the integral of the second term is: 
\begin{equation*}
    \rho e^{-j \theta} S(\omega' - \omega).
\end{equation*}
Finally, the third term differs from the second only by the sign of \(\xi\), so it is equal to \(\rho e^{j \theta} S(\omega' + \omega)\).
\end{proof}

\section{Model Verification}
\label{sec:verification}

In this section we use our mathematical model to study the artifacts generated by Lippmann's procedure and confirm the model predictions experimentally.
We show that even though colors are reproduced relatively accurately, the synthesized spectrum is not the same as the original one, even in a \say{perfect} world, assuming an infinite plate and no windowing due to the development.

Most of the distortions in the spectrum are introduced by the constant term. The most pronounced of them are the oscillations that depend on the bandwidth of \(S\), and thus on the thickness of the plate, see \cref{sec:depth_plate}.

The strength of the oscillations, discussed in \cref{sec:strength}, depends on the development of the plate and the magnitude \(\rho\) of the reflection coefficient \(r\) of the mirror. It is also possible to attenuate the oscillations by changing the photographic process.

Surprisingly, the constant term also causes skewing of the spectrum towards red or blue, depending on the phase \(\theta\) of the reflection coefficient \(r\). We explain this in more detail in \cref{sec:spectrum_skewing}.

Before we delve into details, let us note that our analysis here is mostly qualitative. To perform a quantitative analysis of our model, one would need to measure additional parameters. It is possible to estimate some of them; for example, in \cite{baechler2021shedding} we have estimated the plate thickness \(Z\) from both spectrometric and electron microscopic measurements (Fig. 5B), which resulted in agreement within \SI{10}{\percent}. 
More details on the parameters' estimation are also provided in~\cite{baechler2018sampling}, where an algorithm is proposed to estimate the original spectrum, jointly with the thickness and the development decay.
Using this algorithm, we showed in~\cite{baechler2021shedding} that if the sensitivity of the recording medium is available, the original spectrum can be recovered with acceptable precision, which further confirms the validity of our model.

\subsection{Plate thickness}
\label{sec:depth_plate}

The interference pattern in the gelatin is the generalized Fourier cosine transform of the original spectrum, where lower frequencies are located at the mirror side of the plate. From that perspective, it is clear that the thickness of the plate has a low-pass effect on the synthesized spectrum. This effect has been already observed by Ives~\cite{Ives1908}.

\begin{figure}[!tb]
\centering
	\includegraphics[width=1\linewidth]{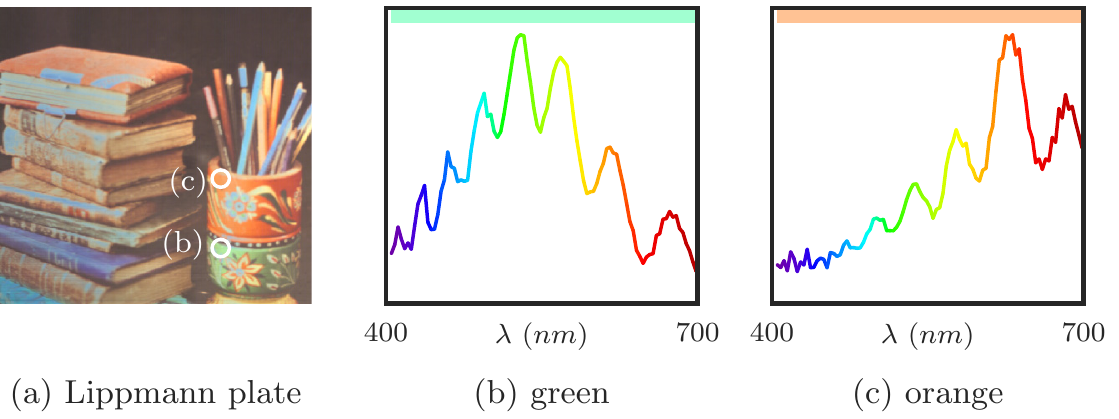}
\caption[]{Hyperspectral acquisition of the Lippmann plate from \cref{fig:lippmann_photos}: (a) close-up of the plate in RGB colors; (b) a green spectrum; (c) a red-orange spectrum. The spatial location of the spectra displayed is shown in (a) and the uniform band on top of each spectrum plots represents the corresponding RGB color.
\label{fig:hyspex} 
}
\end{figure}

Thickness of the plate also controls the frequency of the oscillations in the spectrum (visible in \cref{fig:hyspex}). The main cause of this are the oscillations in the constant term of \(H(\omega, \omega')\) depicted in \cref{fig:h}. Oscillations are also due to the filtering term. As the filter in the spacial domain has a sharp cutoff (due to the sharp end of the plate), it introduces Gibbs ripples. Nevertheless, oscillations due to filtering are not pronounced for smooth real-world spectra. 

As shown in \cref{fig:hyspex}, we performed a hyperspectral acquisition of the plate from \cref{fig:lippmann_photos}.
It is clear that the spectra exhibit the oscillations predicted by our reflection-based theoretical model. 
Furthermore, we can approximate the thickness of the plate from these oscillations. The spectra from \cref{fig:hyspex} present $8$ oscillations in the visible frequency range $\Delta \omega$, so they have a period of $T = \frac{\Delta \omega}{8}$. Comparing this value with the expression for \(S(\omega)\), we deduce that $\frac{2\pi}{T} = \frac{2 Z}{c}$, or $Z \approx 3.7~\si{\mu m}$. This value is in line with the typical thickness of Lippmann plates.

\subsection{Strength of the oscillations}
\label{sec:strength}
Given that the main source of oscillations is the constant term of \(H(\omega, \omega')\), their strength can be modulated by reducing the oscillations in the filter \(S(\omega')\) or by changing the relative strength of the constant and filtered components. 
The former can be achieved during development. The latter can be achieved in three ways: by changing the shape of the input spectrum, by changing the reflectivity \(r\) of the mirror and by further changes to the photographic process. In this section, we describe all these effects.

\subsubsection{Development}
\label{sec:strength_development}

\begin{figure}[!tb]
\centering
	\includegraphics[width=1\linewidth]{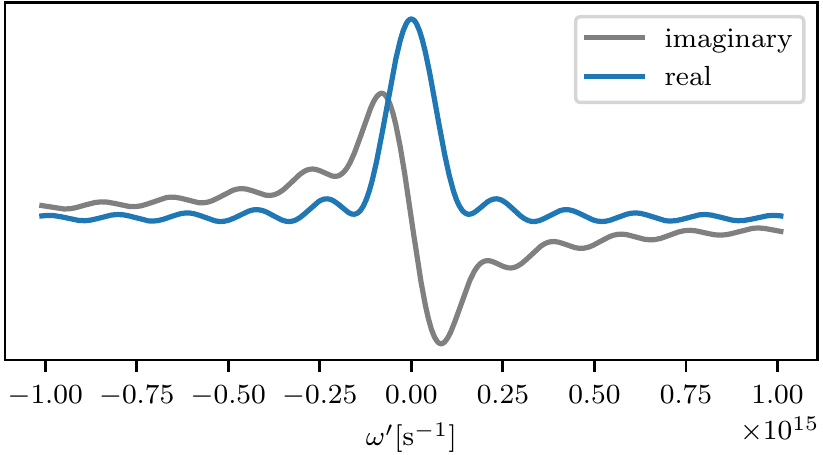}
\caption[]{The filter \(S(\omega')\) for \(Z=5~\si{ \mu} m\) and \(\tau=\SI{3e5}{\per\meter} = 1.5/Z\). Compared to \cref{fig:h}, the oscillations are less pronounced.
\label{fig:h_dev} 
}
\end{figure}

Due to its decaying nature, the development curve attenuates the high frequencies and has a smoothing effect on the oscillations created by the sharp cutoff at the end of the plate, see \cref{fig:h_dev}.

To show the effect of the development, we look directly at the density of developed silver inside a plate.
The variations in the density of metallic silver cannot be observed directly with optical microscopes since the size and spacing of the patterns are beyond the resolving power of visible light.

\begin{figure}[!tb]
\centering
    \begin{tikzpicture}
    \node[inner sep=0pt] (pipeline) at (0,0){
    \includegraphics[width=\linewidth]{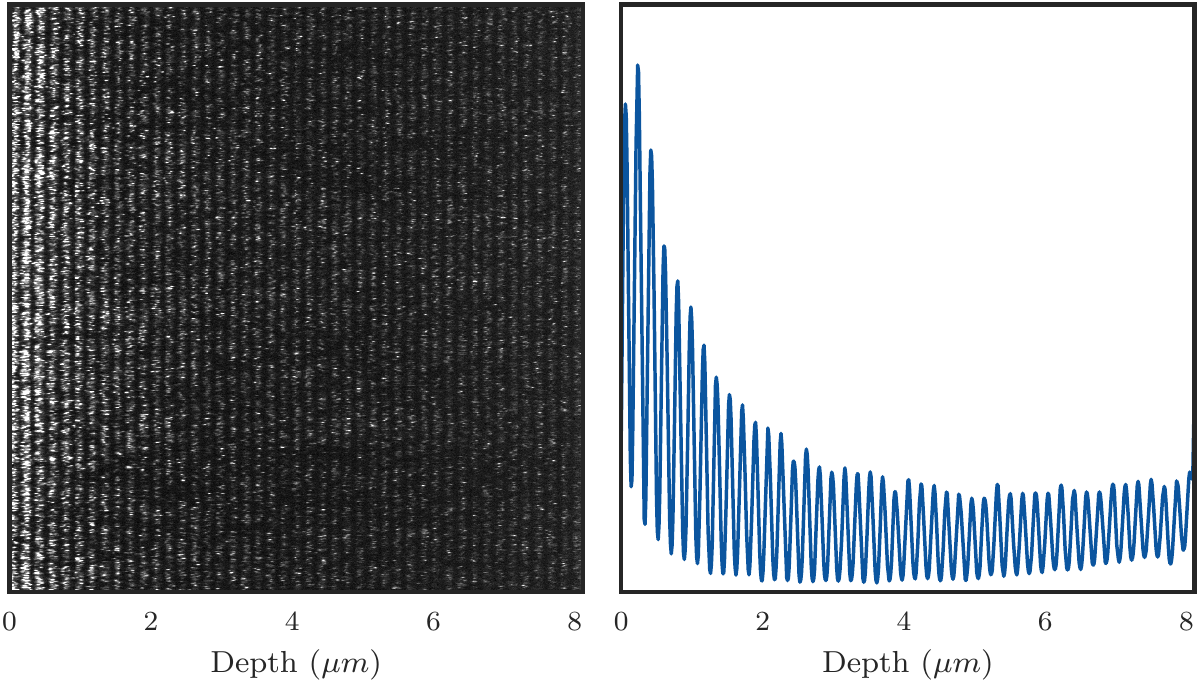}};
    \node[] (a) at (-2.1, -2.8) {(a)};
    \node[] (b) at (2.3, -2.8) {(b)};
    \end{tikzpicture}
\caption[]{Capture of the interference patterns. (a): electron microscope image of a slice of a Lippmann plate created with a $531$~\si{\nm} laser. After fixing, we can observe interference patterns created by the varying density of the metallic silver grains. (b): these patterns are clearly visible after averaging the columns.
\label{fig:fib} 
}
\end{figure}

We use electron microscopy to obtain sub-wavelength resolution imaging.
Electron microscopes cannot distinguish between undeveloped silver halide and metallic silver particles. To see interference patterns, it is necessary to wash out the remaining silver halide; this can be done by fixing the plate.
\Cref{fig:fib} shows an example of the metallic silver particles density resulting from the exposure of a plate to a $531$ \si{nm} laser.

The interference pattern is clearly noticeable in the micrograph from \cref{fig:fib}a as well as when averaging it across pixels located at the same depth in \cref{fig:fib}b. Additionally, we can observe the decay due to development in \cref{fig:fib}b.

\subsubsection{Spectrum shape}
\label{sec:strength_shape}

The strength of the constant term is proportional to the total energy of $P(\omega)$. This is because the mean of the  interference pattern tends to be proportional to the energy of $P(\omega)$; we can see that the interference patterns stabilize around this value. 
The strength of the filtering term at a given frequency \(\omega\) depends mostly on the energy of \(P(\omega)\) in the vicinity of \(\omega\). 

Therefore, for recorded waves with the same peak value of the spectrum, the more the energy of the wave is spread over the visible spectrum, the stronger are the oscillations of the reflected spectrum (relatively to the peak value of the reflected spectrum).

\subsubsection{Reflection coefficient}
\label{sec:strength_reflection}

\begin{figure}[!tb]
\centering
	\includegraphics[width = \linewidth]{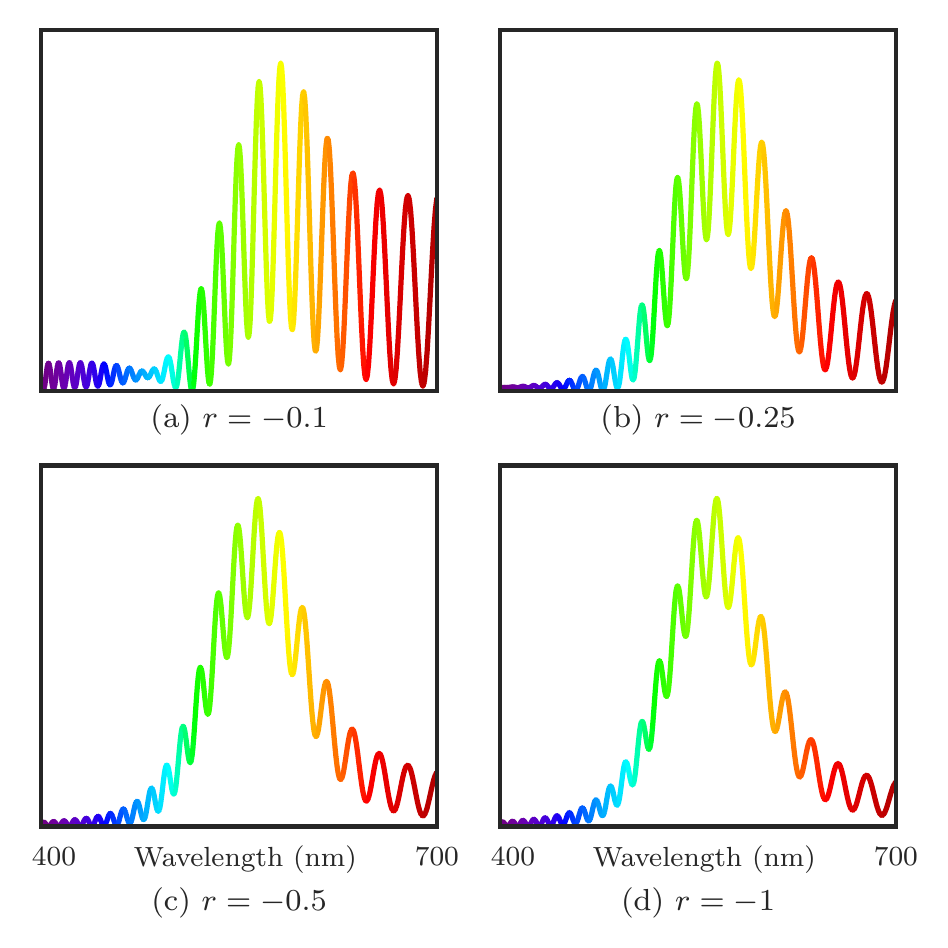}
\caption[]{Reflected spectra based on the Gaussian spectrum for $r = -0.1, -0.25, -0.5, -1$. We observe that the oscillations are proportionally more important in spectra with lower intensities. The spectra have been re-scaled such that they appear similar, but in fact the proportion of reflected light is higher for higher absolute values of $r$.
\label{fig:different_rhos} 
}
\end{figure}

The strength of the oscillations is also influenced by the reflection coefficient. When $\rho < 1$, the interference patterns take the form of partial standing waves. The constant term is multiplied by the value $1+\rho^{2}$ and the filtered term by $\rho$. Subsequently, the constant term becomes relatively stronger as $\rho$ becomes smaller. 
As a result, the spectrum contains stronger oscillations for smaller values of $\rho$. This effect is illustrated in \cref{fig:different_rhos}. 

In Lippmann photography, there are essentially two media that are used to create interference: mercury and air.
For an interface between glass and mercury $\rho = 0.71$ and $\theta = 148^\circ$, and for an interface between glass and air $r = 0.2$, or equivalently, $\rho=0.2$ and $\theta = 0$.
This explains why the oscillations are more intense in plates made with air reflectors than those made with mercury reflectors, which can be observed in \cref{fig:skewing_experiment}.

\subsubsection{Holographic process}
It turns out that the oscillations can mostly be removed from the reflected spectrum by bleaching the plate, as it is often done in holography.
Specifically, a fixation-free rehalogenating bleaching has to be done.

A rehalogenating bleach is a solution containing an oxidizing agent and an alkali metal halide~\cite{Bjelkhagen1993,Hariharan:90}.
When applying it to a developed emulsion, the metallic silver is oxidized to silver ions, which combine with the supplied halogen ions to form tiny silver halide crystals~\cite{Hariharan:90}.
Because of their size, these particles are so soluble in the bleach that they diffuse over a range of the order of one micrometer~\cite{Hariharan:88}.
They end up aggregating onto undeveloped silver halide grains, which are typically much bigger and thus much less soluble~\cite{Hariharan:90}.
So, after the whole processing, silver halide has been essentially redistributed in the emulsion, with a final density following closely the opposite of the interference pattern from exposure~\cite{Bjelkhagen1993,Hariharan:90}.

A bleached emulsion does not contain metallic silver, only transparent silver halide grains.
But these grains have a higher refractive index than gelatin, and changes of refractive index also cause reflection~\cite{Bjelkhagen1993}.\footnote{
This also happens with unfixed, unbleached Lippmann plates.
However, when using the common Lumi\`{e}ere developer, this effect is partly offset by tanning (local hardening from development products, which also increases refractive index but in exposed parts)~\cite{Bjelkhagen1993}.
}
As in an unbleached plate, these reflections at different depths interfere differently for different colors, so a bleached Lippmann plate also reflects light in a selective way.
However, the reflection-based model described in \cref{sec:synthesis} cannot be applied to a bleached plate; we need to model changing index of refraction.

A refraction-based model has been postulated already by Lippmann~\cite{Lippmann1894} and Wiener~\cite{Wiener1899}. It was later refined by Nareid and Pedersen \cite{Nareid1991}, who extended the model to a complex index of refraction. More precisely, they assume that the refractive index $n(z)$ at depth $z$ of the exposed plate could be modified proportionally to the varying part of the intensity of the interference field:
\begin{align}
n(z) = n_0 + \gamma \int_{0 }^{\infty} P(\omega) \rho \cos\left(\frac{2 \omega z}{c } - \theta  \right)d\omega
\end{align}
where \(n_0 > 1\) is the refractive index of the medium after exposition and \(\gamma<0\) is simply a proportionality factor.
The assumptions of this model are reasonable for a plate bleached according to the described method, because the average index of refraction after processing is nearly identical to the one before processing~\cite{Bjelkhagen1993}, and thus spatially constant.

\begin{figure}[!tb]
\centering
	\includegraphics[width=1\linewidth]{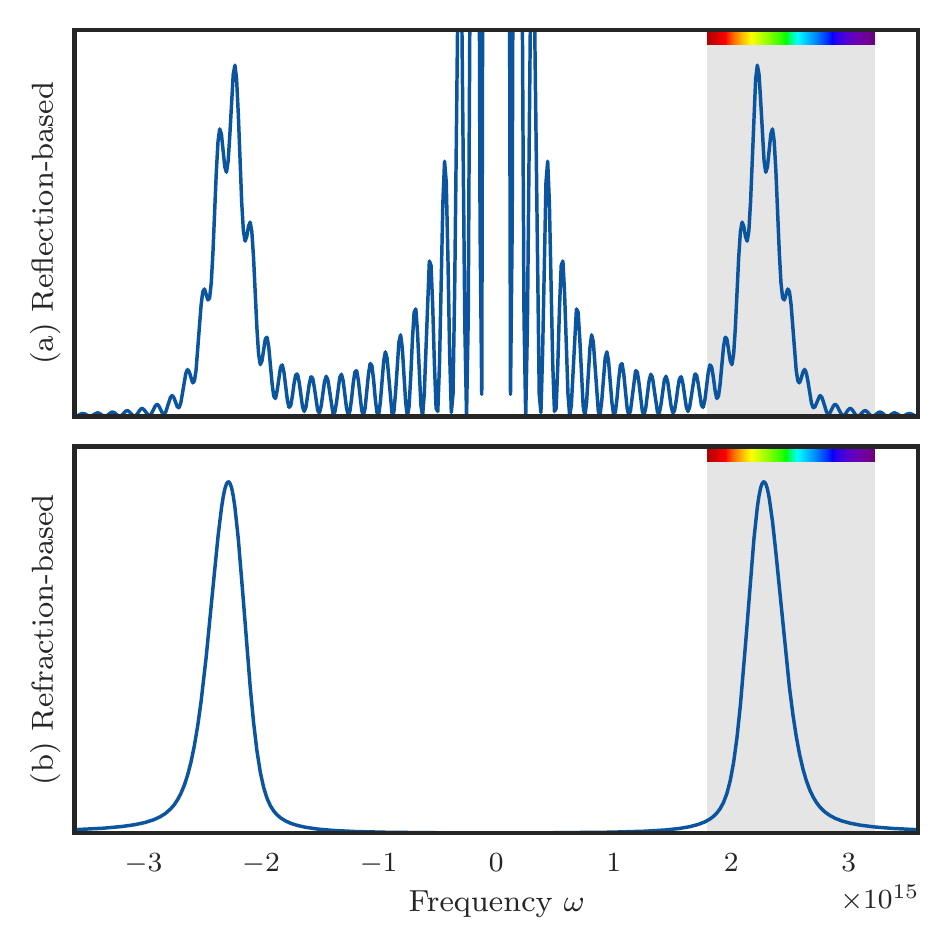}
\caption[]{Simulation of a spectrum reflected by a plate of width $Z= 5~\si{\mu m}, r=-1$, according to (a) the reflection-based model and (b) the refraction-based model.
The original has a Gaussian shape.
\label{fig:reflection_vs_refraction} 
}
\end{figure}

In this model, assuming that surface reflections are removed with a prism of refractive index \(n_0\), the only significant component of the kernel is the filtering term.
The absence of the constant term usually leads to a much smoother reflected spectrum than in the reflection-based model, the only oscillations being the Gibbs ripples due to low-pass filtering. 
Furthermore, this model is not subject to the spectrum-skewing effect described in the next section.
However, spectra are still skewed towards blue (with minimal effect on monochromatic ones): the additional differential operator (reflection comes from changes of refractive index) makes the reflected power spectrum proportional to the squared frequency.
A comparative example of two virtual Lippmann plates from the same exposing spectrum according to the reflection-based and refraction-based models is shown in \cref{fig:reflection_vs_refraction}.

\begin{figure}[!tb]
\centering
    \begin{tikzpicture}
    \node[inner sep=0pt] (pipeline) at (0,0){
    \includegraphics[width=\linewidth]{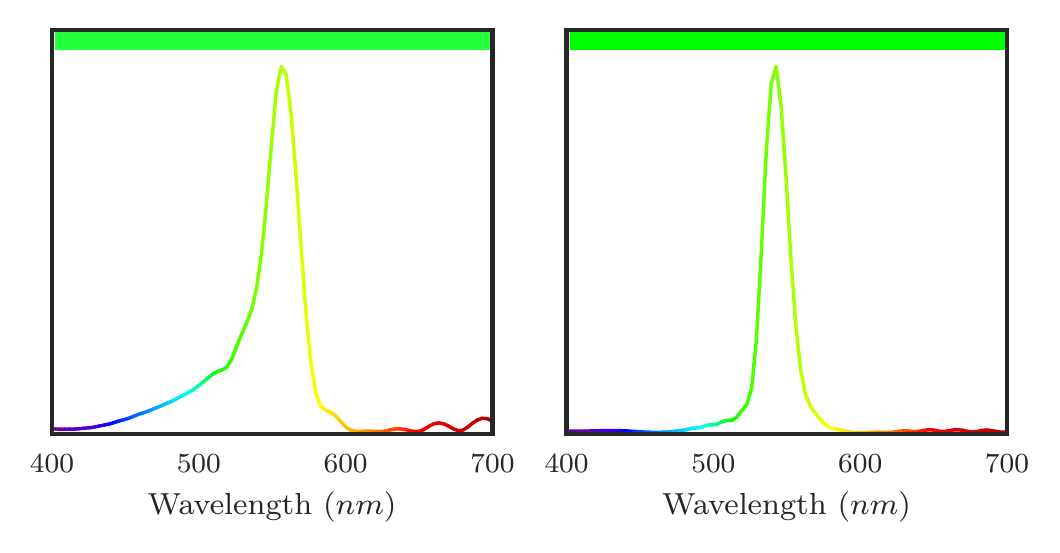}};
    \node[] (a) at (-2.2, -2.4) {(a)};
    \node[] (b) at (2.2, -2.4) {(b)};
    \end{tikzpicture}
\caption[]{Two different development procedures and their effects: (a) the Lumi\`{e}re developer results in spectra that can be explained with the reflection-based model, and (b) holographic recipes produce spectra that can be inferred from the refraction-based model.
\label{fig:lipp_vs_holo} 
}
\end{figure}

The presence or absence of these  effects in a measured plate can help us identify the closest model and consequently the dominant underlying physical process involved.
To illustrate this statement, we created two plates using a $550$ \si{nm} band-pass filter (used later in \cref{fig:skewing_experiment}d). One plate was developed using the historical Lumi\`{e}re developer, which was the typical Lippmann developer in the early 1900s, and the other one was developed and bleached with the method described above. The spectra synthesized by these plates are shown in \cref{fig:lipp_vs_holo}. We notice that the spectrum from \cref{fig:lipp_vs_holo}a clearly exhibits the usual skewing effect as well as moderate oscillations; on the contrary, the spectrum from \cref{fig:lipp_vs_holo}b is relatively symmetric (still with more blue, as expected) and has tiny oscillations (due to inevitable small refractive index mismatch at the surface of the emulsion).

Let us also note that there can be other models that do not contain the constant term, or a weaker one. For example, the constant term can be attenuated in the model by choosing a different approximation of the Hurter and Driffield curve.

\subsection{Spectrum Skewing}
\label{sec:spectrum_skewing}

The constant term of \(H\) has a visible effect not only on the shape of the reflected spectrum, but also on the perceived color. The direction of the change (towards blue or towards red) of the perceived color depends on the phase \(\theta\) of the coefficient of reflection \(r=\rho e^{j\theta}\). In this section we describe this effect and illustrate it by comparing plates created with mercury mirror and with air mirror.

 It is evident that different values of $r$ will generate different interference patterns inside the plate. For example, with $r=1$ a peak is formed at the entrance ($z=0$) of the plate, whereas with $r=-1$, we have a node.

 The filtered component of the reflected spectrum is modulated with $r$, hence it undergoes a rotation in the complex plane that is determined by the value of $\theta$. The constant component is modulated by $1+\rho^{2}$ and is never rotated. This means that these two components interfere with each other differently for different values of $r$.

\begin{figure}[!tb]
\centering
	\includegraphics[width =  1\linewidth]{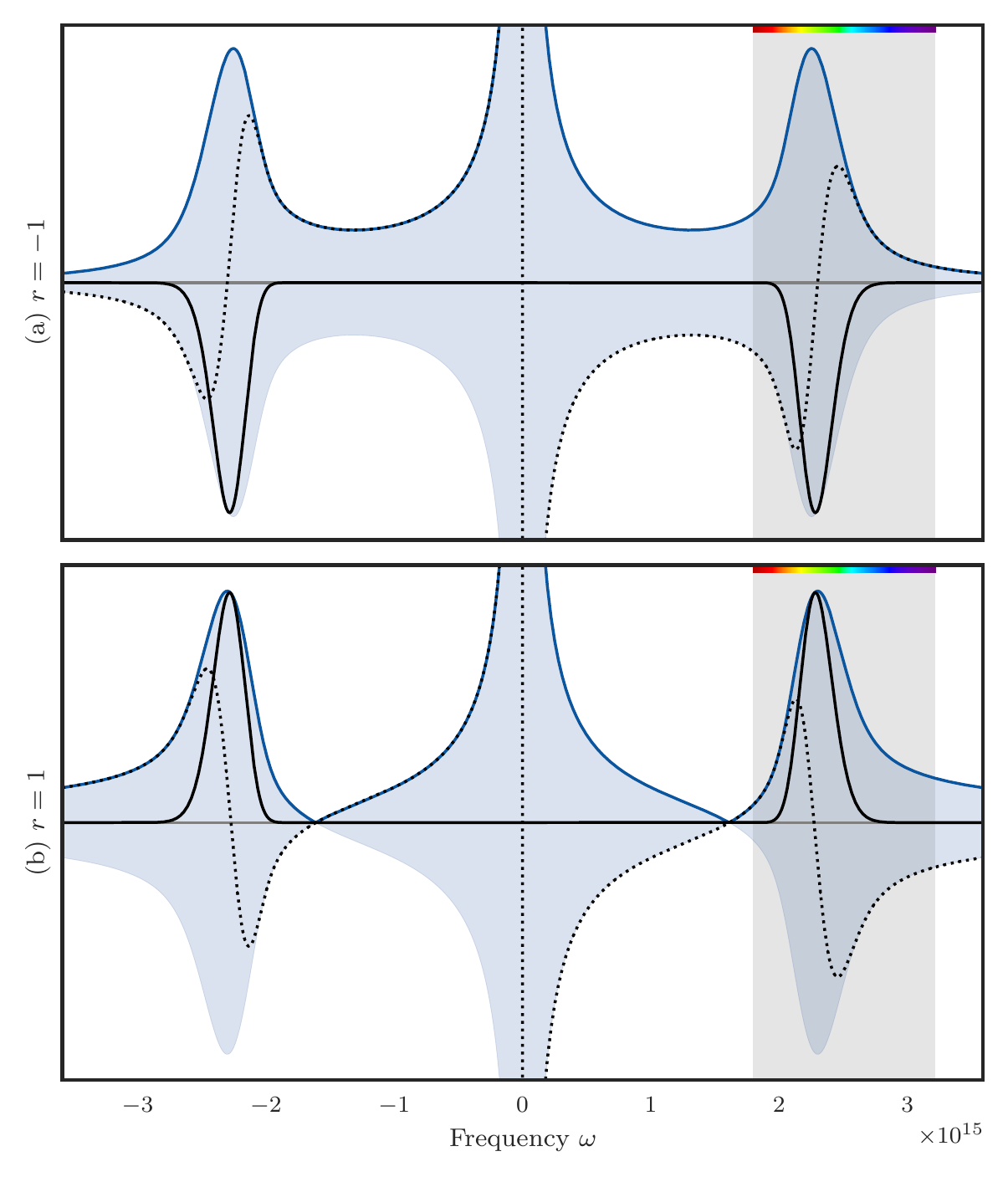}
\caption[]{Skewing effect on the spectrum reflected by an infinitely thick plate that recorded the Gaussian-shaped spectrum for (a) $r = -1$ and (b) $r = 1$. The black lines show the complex wave (the continuous curve is the real part and the dotted curve the imaginary part) and the blue lines show its envelope, which correspond to the absolute value of the amplitude. We observe that (a) the reflected spectrum for $r=-1$ exhibits a negative skew and (b) the reflected spectrum for $r=1$ a positive skew.
\label{fig:skewing_effect} 
}
\end{figure}

\begin{figure}[!tb]
\centering
	\includegraphics[width=1\linewidth]{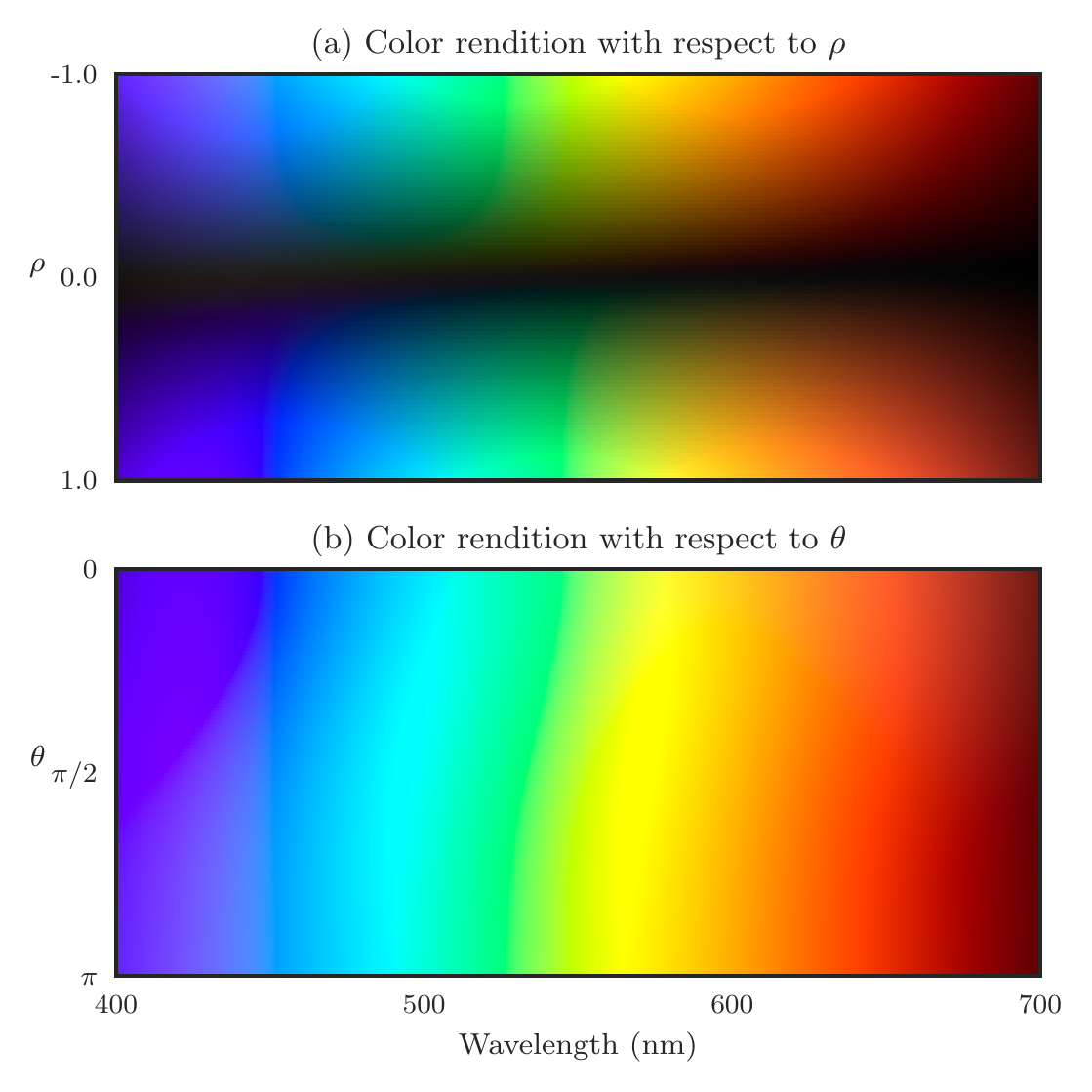}
\caption[]{Skewing effect on the spectrum: Given an original Gaussian spectrum centered at the different visible wavelengths and with $\sigma = \SI{30}{\nano\meter}$, we show the reproduced colors for (a) varying magnitudes $\rho$ (for the phase \(\theta = 0\)) and (b) phases $\theta$ of the reflection coefficient $r$ (for the amplitude \(\rho=1\)).
The spectra are first converted to the CIE 1931 XYZ color space and then to RGB.
\label{fig:spectrum_skewing} 
}
\end{figure}

\Cref{fig:skewing_effect} shows what we call a \emph{skewing} effect on the spectrum.
For $r = -1$, the imaginary part of $S(\omega')$ and $-S(\omega' - \omega)$ add up constructively for red colors and destructively for blue colors, as seen in \cref{fig:skewing_effect}a. On the other hand, for $r=1$, \cref{fig:skewing_effect}b shows the opposite phenomenon, namely the spectrum is skewed such that it favors blue colors over red colors.
The direction of the skewing is independent of the thickness of the plate and is also present in the simulations of plates of infinite thickness.  

In practice, when the reflection is due to the interface with air, we have $r=0.2$, so we should expect stronger blues, to the detriment of reds since $|\theta| < \pi/2$. With mercury we have $\rho = 0.71$ and $\theta = 148^\circ$ and therefore, red colors should dominate since $pi \geq |\theta| > \pi/2$. Given a Gaussian spectrum centered around a specific wavelength, \cref{fig:spectrum_skewing} illustrates the expected reproduced colors for different values of $\rho$ and $\theta$. It suggests that bright reds are more challenging to reproduce with air, and blues are harder to represent with mercury. We also notice a shift of the reflected spectra: for instance, greens should appear blueish with air and yellowish with mercury. Moreover, a smaller value $|\rho|$ leads to darker colors; this explains why, in general, plates made with mercury appear brighter than those made with air.

\begin{figure}[!tb]
\centering
	\includegraphics[width = \linewidth]{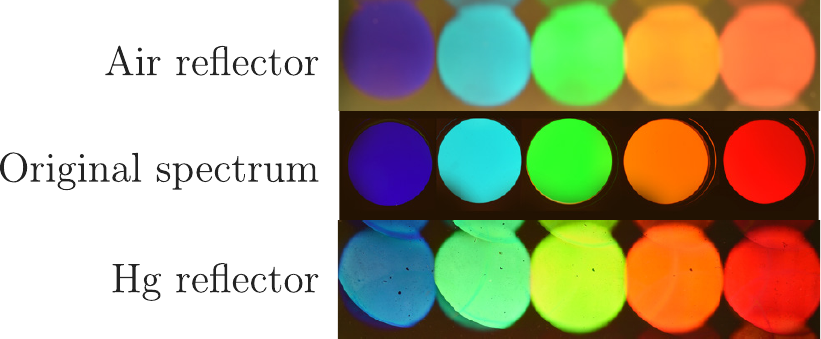}
\caption[]{Skewing effect with air and mercury mirror on narrow Gaussian-shaped spectra: the middle row shows the five filters used to expose the plates, while the top and bottom rows show the resulting plates after development.}
\label{fig:skewing_suggestion} 
\end{figure}

\begin{figure}[!tb]
\centering
	\includegraphics[width = \linewidth]{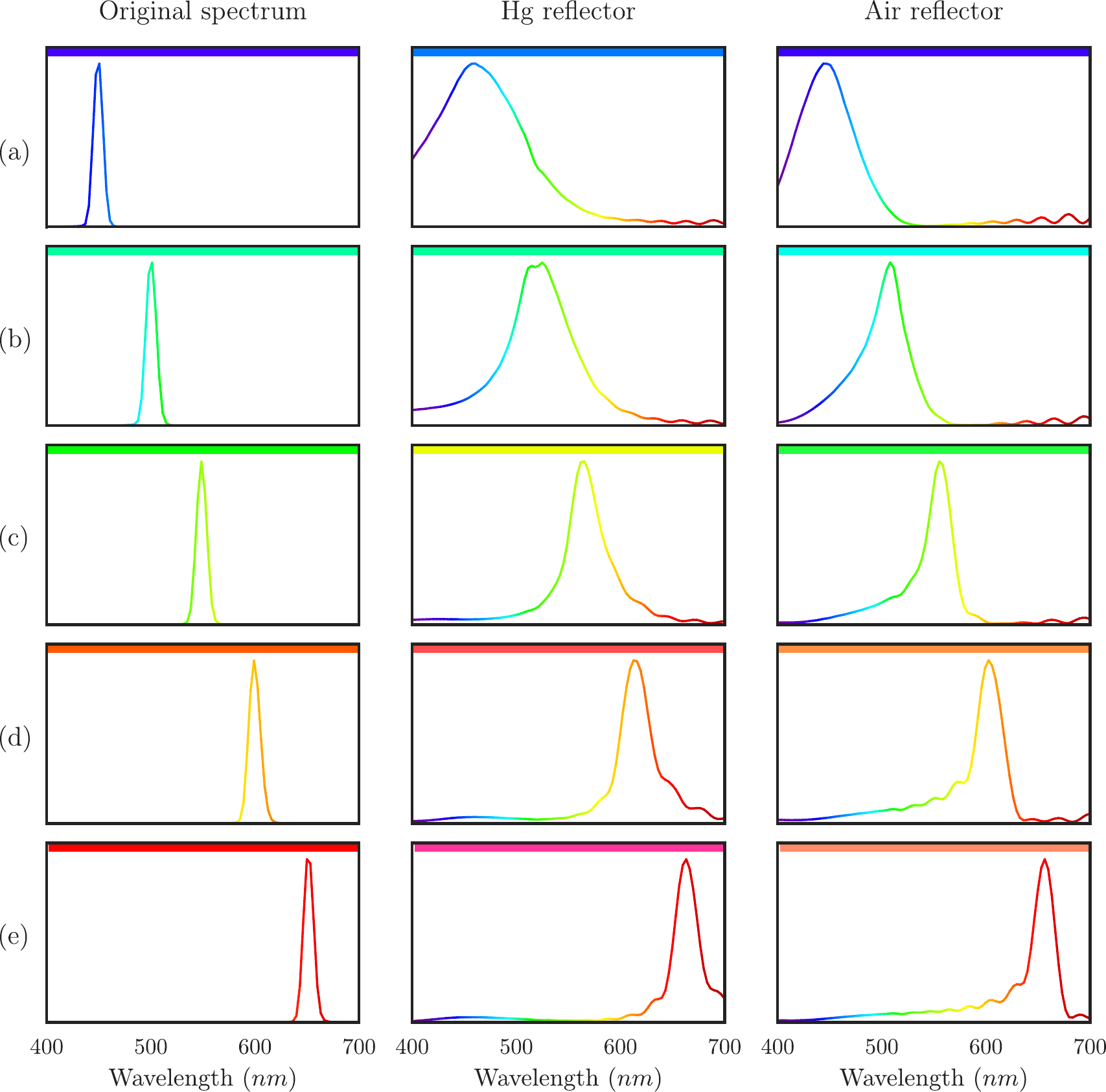}
\caption[]{Skewing effect with air and mercury mirror on Gaussian-shaped spectra from \cref{fig:skewing_suggestion}: measured spectra corresponding to each of these five bands: (a) $450$ \si{nm}, (b) $500$ \si{nm}, (c) $550$ \si{nm}, (d) $600$ \si{nm}, and (e) $650$ \si{nm}.
\label{fig:skewing_experiment} 
}
\end{figure}

In our previous work \cite{baechler2021shedding} we have shown an experiment confirming the results from \cref{fig:spectrum_skewing}. Here we run a more detailed experiment by placing Gaussian-shaped bandpass filters having a bandwidth of approximately $10$ \si{nm} (\cref{fig:skewing_suggestion}, middle) in front of a halogen light source. As before, we create two plates: one with mercury and one with air as reflectors.
Photographs of the resulting plates are shown in \cref{fig:skewing_suggestion}. For this experiment, we also acquired hyperspectral measurements of the original spectra as well as the spectra reflected by these plates (see \cref{fig:skewing_experiment}).
We see that the reproduced colors are in agreement with our predictions. Indeed, the spectra of the plates realized with a mercury reflector exhibit a  skew towards red while those made with an air reflector skew towards blue.

We can also observe that the oscillations are not significant, which suggests that they are damped by a development curve strongly decaying with depth (\cref{sec:strength_development}).
Furthermore, their amplitude is slightly larger in the plates made with air, as predicted by \cref{sec:strength_reflection}.

As a final remark, the strength of the skewing depends on the shape of the spectrum, in the same way as the strength of oscillations does (\cref{sec:strength_shape}).  In the extreme case where the original spectrum is a Dirac delta (monochromatic light), the tail of the center component has a weak influence on the visible spectrum, especially for thick plates.

\subsection{Materials and Methods}

We used U08C photographic plates from Ultimate Holography.
Our plates were developed using a pyrogallol-ammonia developer and were not fixed.

All spectral measurements were captured using a FLAME-S-XR1-ES spectrometer from Ocean Optics (spectral resolution \SI{\sim2}{nm}, 2048 bands) and some custom focusing optics, such that the measured area is less than \SI{1}{\milli\meter^2}.

Electron microscopy was done by EPFL's Interdisciplinary Centre for Electron Microscopy using a scanning electron microscope equipped with a focused ion beam for sample preparation.

Color photographs reproduced here were taken with a Nikon D810 color-calibrated digital single-lens reflex camera.

\section{Conclusion}

A thorough inspection of Lippmann's Nobel prize winning paper~\cite{Lippmann1894} enabled us to elucidate a misconception that went unnoticed for more than a century regarding the perfect reconstruction of the synthesized spectrum with infinite plates. Furthermore, our formulation provided a clear explanation of the color reproduction of Lippmann plates. This allowed us to characterize and describe various effects such as the role of low-pass filter played by the thickness of the plate, the skewing effect induced by the choice of reflector, and the presence of oscillations in the synthesized spectra. Finally, we have qualitatively confirmed the predictions of our model experimentally.

\section{Acknowledgements}

The realization of our own Lippmann plates would not have been possible without the help and contributions of the following people: Filipe Alves, who was the first to show us that Lippmann photography can be made practical and who shared some of his secrets on how to realize beautiful plates; Yann Pierson, who assisted us with our chemistry shopping list and lent us his chemistry laboratory; and the GR-CEL group and in particular Prof. De Alencastro and Sylvain Coudret, who let us use their chemistry laboratory and assisted us with our chemistry experiments.
We are also grateful to Prof. Jean-Marc Fournier for sharing his infinite knowledge of the Lippmann process and to Yves Gentet, for his advice on the development and making of holographic plates.


\bibliographystyle{IEEEtran}
\bibliography{bib}

\end{document}